\documentclass[twocolumn,showpacs,preprintnumbers,amsmath,amssymb,prb]{revtex4}
\usepackage{graphicx}

\begin{document}

\title{Elastic constants and volume changes associated with two
high-pressure rhombohedral phase transformations in vanadium}

\author{Byeongchan Lee}
\author{Robert E. Rudd}
 \email{robert.rudd@llnl.gov}
\author{John E. Klepeis}
\author{Richard Becker}

\affiliation{Lawrence Livermore National Laboratory, Livermore, California 94551}

\date{\today}

\begin{abstract}
We present results from {\em ab initio} calculations of the
mechanical properties of the rhombohedral phase ($\beta$) of vanadium metal reported in
recent experiments, and other predicted high-pressure phases
($\gamma$ and bcc), focusing on properties relevant to dynamic experiments.
We find that the volume change associated with these transitions is
small: no more than 0.15\% (for $\beta$ -- $\gamma$). 
Calculations of the single crystal
and polycrystal elastic moduli (stress-strain coefficients) 
reveal a remarkably small discontinuity in the shear modulus and
other elastic properties
across the phase transitions even at zero temperature where the
transitions are first order.
\end{abstract}

\maketitle

\section{\label{sec:level1}Introduction}

The existence of a high-pressure rhombohedral phase of pure
crystalline vanadium has been the focus of an intense research effort recently.
The first indication of a phase transition came from the theoretical
observation that the C$_{44}$ shear modulus of bcc vanadium decreases and
%diminishes under pressure,
%Other elemental bcc metals (Ta, Nb) have similar features in the pressure dependence of the shear modulus, but in the case of vanadium, the bcc C$_{44}$ is predicted not only to plateau but to
becomes negative at pressures greater than $\sim$1.3~Mbar,\cite{Suzuki,Landa} pressures that
are experimentally accessible.
A negative shear modulus means the
material is mechanically unstable under trigonal (prismatic) shear,
suggesting a phase transition.  At that time the experimental evidence
showed no phase transition up to 1.54~Mbar.\cite{Kenichi}
Then recently,
Mao and coworkers \cite{Mao} conducted x-ray diffraction experiments in
the diamond anvil cell (DAC) up to 1.5~Mbar and found features in the
diffraction peaks that were consistent with a second-order phase
transformation to a rhombohedral structure with an R$\bar{3}$m point group
symmetry at pressures above 0.69~Mbar. It was soon confirmed that
density functional theory (DFT) finds the rhombohedral phase to be
the ground state at zero temperature and pressures above 0.8~Mbar, in
reasonably good agreement with experiment.\cite{Lee}
In fact, it was shown that
DFT predicts additional phase transformations that had not been found in experiment, i.e. a first-order transformation to a different rhombohedral
structure at 1.2~Mbar and a third transformation back to the bcc
structure at 2.8~Mbar.\cite{Lee}
%These pressures have not been reached in the recent DAC experiments.
The prediction of the existence of the two high pressure phase
transformations has been subsequently confirmed with
DFT phonon calculations.\cite{Luo}

Alternative techniques may provide the pressures
needed to observe the second rhombohedral phase and the
reentrant bcc phase.
Dynamic experiments do not rely on the mechanical integrity of
anvils and are able to reach multi-megabar pressures.
They have been used to study similar transformations
%\cite{Meyers}
such as the diffusionless $\alpha$-$\varepsilon$ transition in iron.\cite{DuvallGraham}
There are several challenges specific to vanadium, however.  The softening
of the shear modulus and the rhombohedral phase transition are
related to subtle electronic effects,\cite{Landa} which are
likely to be weakened by increased temperature.
%Dynamic experiments that generate shocks cause substantial temperature rises.
Recent ramp wave techniques based on Z,\cite{Asay}
laser\cite{Remington,Smith} and graded-density impactor\cite{Nguyen} drives
are able to generate high pressure without the entropy generation
of shockwave techniques, and are therefore preferable in the present context.
Another challenge is that the subtle rhombohedral distortion
($<1^{\circ}$) detected by x-ray diffraction in the DAC is
probably too small for in-situ x-ray diffraction
in dynamic experiments.\cite{Kalantar}
%\cite{Wark,Kalantar}
Indirect techniques are an alternative to detect the
transition.  For example, VISAR free-surface velocity measurements
can detect changes in the density due to a volume change, and
they can be used to infer the longitudinal stress, and hence the
change in
strength if the equation of state is known.\cite{Gupta,Smith}
Rayleigh-Taylor growth rate is another way to probe strength.\cite{Remington}

In this article we use DFT to make predictions about the
properties of high-pressure vanadium relevant to dynamic experiments.
We compute the magnitude of the
volume change associated with the three phase transitions related to
the rhombohedral structure in Section~\ref{sec:level3}.  We also
compute the elastic properties and calculate bounds on, and an
explicit estimate of, the
polycrystalline shear modulus in Section~\ref{sec:level4} and Section~\ref{sec:level5} respectively.  Since the strength is typically
assumed to vary with the shear modulus,\cite{SteinbergGuinan} any
anomalies in the shear modulus are likely to provide a signature in
the VISAR trace.  Indeed, an important motivation for the present
work is to assess whether the bcc shear modulus $C_{44}$ going to
zero is likely to produce a strong signature.
The shear modulus also affects defect energetics, and may have a
measurable effect on transition kinetics.
We consider the implications of our results for dynamic experiments to
detect the high-pressure phases.

\section{\label{sec:level2}Theoretical Background}

The rhombohedral crystal structure of vanadium at high pressure results from a slight
distortion of the bcc structure. Specifically the distortion is
a uniaxial strain along $\langle 111 \rangle$, which remains a
three-fold symmetry axis of the crystal.  This structure is known as the
$\beta$-Po structure ({\em Strukturbericht} A$_{\mathrm{i}}$, Pearson hR1).
% Consider an atom in the bcc structure and three of its nearest neighbors
% arrayed symmetrically about a three-fold crystal axis
% ($\langle$111$\rangle$ in the bcc crystal).  The angle formed between any
% two of these neighbors is arccos(-1/3)=109.47$^\circ$.
% Now imagine the crystal strained uniaxially in the direction
% of this three-fold axis such that these vectors are
% pushed symmetrically in (out) so that the angles between the
% pairs decrease (increase) remaining equal.
% The resulting crystal retains the three-fold symmetry axis $\langle 111 \rangle$.
% It is the rhombohedral $\beta$-Po structure.
It still has a single atom per unit cell, so the rhombohedral transition
may be expected to be diffusionless (martensitic) and likely rapid despite
the small energy difference.
%R$\bar{3}$m, and it may be described by the order parameter $Q = (\alpha -\alpha _0)/\alpha _0$.\cite{Mao}
There are four independent three-fold axes,
so there are four variants of the rhombohedral crystal that
are degenerate in energy.

The ground state of the single-crystal rhombohedral phase has been
determined from first principles using a volume-conserving rhombohedral shear path,\cite{Lee}
%(a generalization of the conventional \cite{Dacorogna} monoclinic distortion).
%
\begin{equation}
T(\delta)=\left(
\begin{array}{ccc}
k & \delta & \delta\\
\delta & k &\delta\\
\delta & \delta & k\\
\end{array}
\right)
\label{eq:transformation}
\end{equation}
in the usual bcc crystal frame
where $k$ is determined from the real positive solution of $\det(T)=1$ to
insure constant volume.
The approach
%here as in our previous calculations
is to use
DFT in combination with a gradient-corrected exchange and correlation
energy functional \cite{GGA} as implemented in the
Vienna Ab-initio Simulation Package (VASP) code along with the projector
augmented-wave (PAW) method.\cite{PAW}
% and standard computational parameters.\cite{Calculation}
%\bibitem{Calculation}
Specifically, the PAW potentials with 13 valence electrons
(3$s$, 3$p$, 3$d$, and 4$s$ states) are used.
The planewave cutoff energy is 66.15~Ry and an unshifted
60$\times$60$\times$60 uniform mesh is used for the $k$-point
sampling: this results in 5216 and 18941 $k$ points in the irreducible Brillouin zone of the unstrained bcc and rhombohedral lattices, and up to 18941 and 54932 $k$ points for the strained bcc and rhombohedral lattices respectively. For all of the calculations we use a primitive cell.

\section{\label{sec:level3}Volume change due to transformation}

In Ref.~\onlinecite{Lee},
we calculated the enthalpy and pressure as functions of strain
along the rhombohedral deformation path,
and used the enthalpy to find any stable or metastable crystal structures.
We noted that the equations of state (EOS) for the bcc and rhombohedral
structures are nearly identical, so their bulk moduli are essentially equal
(differing by no more than 3\%), and reported the EOS of the ground state
up to 2 Mbar.
We now use those data together with additional data on the EOS of the metastable
structures to calculate the volume change associated with
the phase transformations in a readily accessible form.
%Suppose the pressure change
%in going from phase 1 to phase 2 at constant volume is
%$\Delta P_{12}=P_2-P_1$
%and the bulk modulus of phase 2 is $K$.  Then the volumetric strain
%induced by changing the pressure in phase 2 to $P_1$ is
%$\Delta V/V = \Delta P_{12} / K$.
%This relative volume change is plotted in Fig.~\ref{fig:vol}.

\begin{figure}
\includegraphics[width=0.43\textwidth]{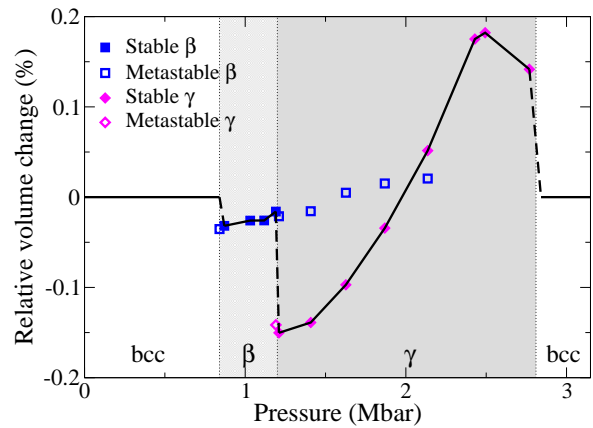}
\caption{(Color online) The volumetric strain $\Delta V/V$ of the ground state
of single-crystal vanadium at zero temperature with respect
to that of the bcc structure. Dashed lines correspond to three volume changes associated with three phase transitions.
The light (dark) gray area represents the pressure range in which
the $\beta$ ($\gamma$) phase is the ground state respectively. The bcc
phase is stable in the other regions.
\label{fig:vol}}
\end{figure}

Using the EOS $P_i(V)$ for the stable and metastable structures 
($i={\mathrm bcc}, \beta, \gamma$), 
we have calculated the associated volume change $\Delta V$ according to 
\begin{equation}
P_j(V_i+\Delta V)=P_i( V_i)
\label{Eq:equalP}
\end{equation}
for pairs of structures $i$ and $j$.
In practice, we have calculated the pressure at a set of volumes
and used piece-wise quadratic interpolatation to solve the equal pressure 
condition (\ref{Eq:equalP}) between those points, equivalent to
the common tangent construction at the phase boundaries.
The EOS data are tabulated in Table~\ref{tab:EOS}.
The relative volume change with respect to the bcc phase, 
$\Delta V/V_{{\mathrm{bcc}}}$, is plotted in Fig.~\ref{fig:vol}.

\begin{table}
\caption{\label{tab:EOS}Equations of state for the bcc,  $\beta$ and $\gamma$ 
phases and metastable structures. Pressures are in Mbar, and volumes are in 
units of the ambient volume $V_o$=13.518~\AA$^3$.}
\begin{ruledtabular}
\begin{tabular}{cccc}
Volume &  \multicolumn{3}{c}{Pressure} \\
		& bcc & $\beta$ & $\gamma$ \\
\hline
1.000 & 0.000 & - & - \\
0.831 & 0.479	& - & - \\
0.804 & 0.596 & - & - \\
0.779 & 0.726 & 0.724 & - \\
0.759 & 0.840 & 0.838 & - \\
0.754 & 0.870 & 0.869 & - \\
0.729 & 1.031 & 1.030 & - \\
0.717 & 1.118 & 1.117	 & 1.110 \\
0.707 & 1.191 & 1.190 & 1.183 \\
0.705 & 1.210 & 1.209 & 1.202 \\
0.681 & 1.408 & 1.407 & 1.399 \\
0.659 & 1.627 & 1.627 & 1.620 \\
0.636 & 1.869 & 1.870 & 1.866 \\
0.614 & 2.136 & 2.138 & 2.140 \\
0.593 & 2.430 & 2.433 & 2.445 \\
0.588 & 2.494 & - & 2.510 \\
0.572 & 2.769	& - & 2.782 \\
0.568 & 2.841 & - & 2.852 \\
0.551 & 3.149 & - & - \\
\end{tabular}
\end{ruledtabular}
\end{table}

% The EOS for a stable or metastable rhombohedral phase defined by $\delta$, bcc if $\delta=0$, has been interpolated from all the calculated $P(\delta, V)$ data due to Eq.~\ref{eq:transformation}. Then the volume change against the bcc phase at given volume ($V_{bcc}$), hence given pressure ($P_{bcc}$), is obtained from the following condition
% %
% \begin{equation}
% P(\delta, V_{bcc}+\Delta V)=P_{bcc}(0, V_{bcc}),
% \end{equation}
% %
% which is in principle identical to the common tangent method used with enthalpy curves. But in reality, the common tangent method involves an enthalpy expansion\cite{Lee} from our constant-volume energy calculations as well as a fit to an available EOS, and hence, prone to more numerical errors. This relative volume change with respect to the bcc phase, $\Delta V/V_{bcc}$, is plotted in Fig.~\ref{fig:vol}.

% FIG

In principle, there is a volume change during the
transformation from bcc ($\alpha$) to the first rhombohedral phase
($\beta$), and a second volume change associated with
the transformation to the second rhombohedral
phase ($\gamma$), and a third volume
change associated with the transformation back to bcc ($\alpha$) at
high pressure.  There is no path connecting $\beta$ and $\gamma$
that preserves the point group, so they must be distinct phases;
the two bcc regions appear to be connected at finite temperature.\cite{Landa}
In practice, the kinetics of the
transformation may cause the transformation to be
overdriven so that the initial phase is retained
in a metastable state past the phase boundary until
the new phase has a chance to nucleate and grow.
For this reason it is interesting to examine the
entire curve in Fig.~\ref{fig:vol}, and not just
the values in between the equilibrium phase boundaries.

In each case the initial $\Delta V$ is a volume change, so the volume is reduced
following the transition.
The volume change associated with the bcc to $\beta$ transformation
is small, 0.03\% or less in magnitude.  It would not be easy to detect such
a small change in a dynamic experiment.  The magnitude of the volume change
associated with the second transformation is larger:
about 0.15\% for the $\beta$ to $\gamma$ transformation at 1.2~Mbar.
The volume change would be about the same if bcc were retained
to a pressure of $\sim$1.2~Mbar and then transformed directly
to $\gamma$.  This
%$\Delta V/V$ is not as great as in the $\alpha$-$\varepsilon$ transition in
%iron, but it may be large enough to detect.
However, if the bcc or $\beta$ phase persists to higher pressures, the volume change becomes progressively smaller and eventually changes sign, becoming a volume expansion near 2~Mbar. The final transition
back to bcc again has a change of over 0.1\% in magnitude.
So the $\beta$ to $\gamma$ transformation has the strongest signature
in terms of volume change, but it may not be large enough to detect.

\section{\label{sec:level4}Single crystal elastic moduli}

% FIG
\begin{figure}
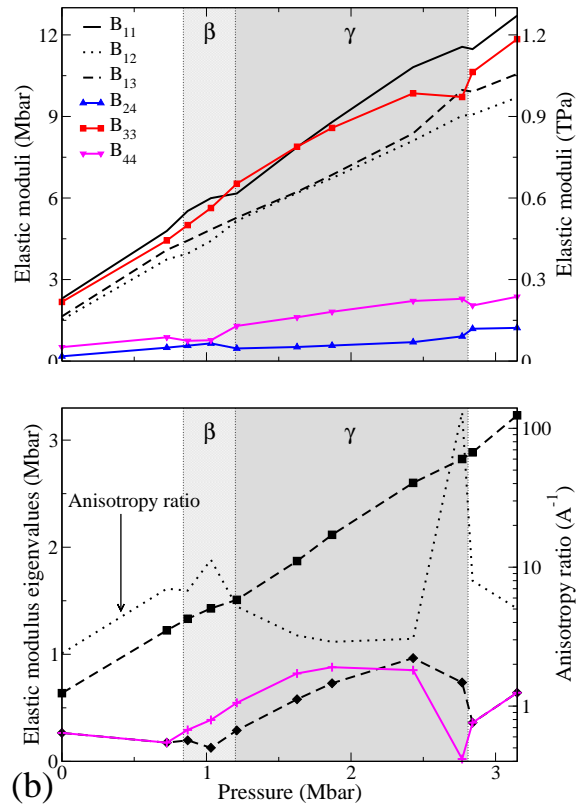

\includegraphics[width=0.42\textwidth]{Bijkl}
\includegraphics[width=0.42\textwidth]{eigenval}
\caption{(Color online) The elastic moduli (specifically, the
stress-strain coefficients)
of the ground state single-crystal structure
%--bcc, $\beta$, $\gamma$, and reentrant bcc--
as a function of pressure;
(a)  $B_{ijkl}$ written in Voigt notation in the frame of the
primitive rhombohedral cell, and
(b) the corresponding eigenvalues of the $9\times9$ stress-strain
coefficient matrix $B_{ijkl}$ (see text).
In the rhombohedral
phase there are six independent elastic moduli (stress-strain coefficients)
(vs.\ three for bcc),\cite{Wallace} and
three independent shear eigenvalues for vanadium metal
(dashed curves represent doubly degenerate eigenvalues).
%Remarkably, both the ground-state bulk modulus (the top curve)
%and the $B'$ (the second curve from the top) are
%fairly smooth through the rhombohedral phase.
The anisotropy ratio $A^{-1}$ is also plotted, showing an extreme
value just below the $\gamma$--bcc transition.
\label{fig:eConst}}
\end{figure}

We next consider how the single crystal elastic moduli change
with pressure.
%The elastic constants of the bcc structure as
%a function of pressure have been calculated previously,\cite{Landa}
%but those of the rhombohedral structure have not.
Specifically, we calculate
$B_{ijkl}(P)$, the elastic moduli with respect to a shear-stress-free
reference state at pressure $P$ (either the bcc or rhombohedral structure,
as specified).  The $B_{ijkl}$ are often called the stress-strain 
coefficients, as we do below.\cite{Wallace} They are directly 
related to sound velocities at high pressure.
%They are equal to $C_{ijkl}$ when the pressure vanishes;
%for $P\ne0$, the relation is\cite{Wallace}
%\begin{equation}
%B_{ijkl}= -P(\delta_{jl}\delta_{ik}+\delta_{il}\delta_{jk}-\delta_{ij}\delta_{kl}) + C_{ijkl} \;.
%\label{eq:BvsC}
%\end{equation}
%
%$B_{ijkl}$ can be obtained by expanding the formulas for
%$C_{ijkl}$  (for example Eq.\ (1) of Ref.~\onlinecite{Ravindran}) using the relation\cite{Wallace}
%\begin{equation}
%B_{ijkl}= -P(\delta_{jl}\delta_{ik}+\delta_{il}\delta_{jk}-\delta_{ij}\delta_{kl}) + C_{ijkl} \;.
%\label{eq:BvsC}
%\end{equation}
$B_{ijkl}$ can be obtained from the deformation paths used for orthorhombic lattice\cite{Ravindran} or for trigonal lattice,\cite{Zhao} and the details are given in the appendix.

The six independent stress-strain coefficients,
$B_{11}$,  $B_{33}$,  $B_{12}$,  $B_{13}$,  $B_{44}$,  and
$B_{24}$ (here given in Voigt notation in the
rhombohedral frame with
%the three-fold axis in direction~3:
directions $[\bar{1}10]_{bcc}$, $[\bar{1}\bar{1}2]_{bcc}$ and $[111]_{bcc}$ as 1, 2 and 3, respectively),
are plotted in Fig.~\ref{fig:eConst}(a). They
are discontinuous at the first-order phase transitions;
however, within the
domain of each stable phase, most of the
stress-strain coefficients increase monotonically with pressure. The exception is $B_{33}$
 near the $\gamma$ to reentrant bcc phase boundary (roughly 2.8~Mbar). Since $B_{33}$ is associated
with uniaxial strain along the three-fold axis, its
anomalous behavior is suggestive, but a better presentation is needed
to separate the effects of shear and compression.  We turn to it now.

% $B_{33}$ is the uniaxial
% elastic constant in the three-fold axis characteristic of the rhombohedral
% lattice, which by definition gives the curvature of the energy along
% the rhombohedral shear path.
% % described by Eq.\ (\ref{eq:transformation}).
% The complex behavior of $B_{33}$ on top
% of the abrupt change in the slope leads to a nontrivial variation in the
% crystalline anisotropy (RER: quantify).
% The other elastic constant that is associated with the
% shear path, $B_{44}$, also shows weak softening around the $\gamma$-to-reentrant
% bcc transition.

Indeed, there is a remarkable approximate continuity of the elastic
properties across the phase transitions that is not readily apparent
from the elements of $B_{ijkl}$.
The bulk modulus of the rhombohedral
phases is within 3\% of that of the bcc structure, as we already
discussed.
%
% We define the fourth-rank tensor $J_{ijkl}$ as
% \begin{equation}
% J_{ijkl} = \frac{1}{2} B_{ijmn} \left( \delta _{mm'} \delta _{nn'} - \frac{1}{3}  \delta _{mn} \delta _{m'n'}\right) B_{klm'n'}
% \end{equation}
% so that the second invariant of the deviatoric stress is given
% by $J_2 = J_{ijkl} \varepsilon _{ij} \varepsilon _{kl}$, and
% the von Mises stress is $\sqrt{J_2}$.
% The eigenvalues of the $9\times9$ matrix $J_{(ij)(kl)}$
% provide a description of the elasticity that is less coordinate dependent.
%
The eigenvalues of the $9\times9$ matrix $B_{(ij)(kl)}$
provide a description of the elasticity that is less coordinate
dependent, but there is a technical issue.  In the rhombohedral phase,
shear and compression are mixed in the sense that a non-equiaxed strain
is required to produce purely hydrostatic pressure, and the tetragonal
strains to produce hydrostatic pressure and pure shear are not
orthogonal.  To eliminate any ambiguity, we restrict to the space
of constant volume strains using a projection matrix
$\Pi_{(ij)(kl)}=\delta _{(ij)(kl)} - \frac{1}{3} \delta _{ij}\delta _{kl}$.
Then the $9\times9$ matrix $\Pi B \Pi /2$ has 5 nontrivial eigenvalues,
corresponding to different shear moduli. This matrix is closely
related to von Mises stresses.
%
%The 6 nontrivial eigenvalues are
% \begin{widetext}
% \begin{eqnarray}
% 3K & = &
% \frac{{B_{11}} + {B_{12}} + {B_{33}} +
%      {\sqrt{{{B_{11}}}^2 +
%          2\,{B_{11}}\,{B_{12}} +
%          {{B_{12}}}^2 + 8\,{{B_{13}}}^2 -
%          2\,{B_{11}}\,{B_{33}} -
%          2\,{B_{12}}\,{B_{33}} +
%          {{B_{33}}}^2}}}{2} \\
% E' (1,2) & = &
%   \frac{{B_{11}} - {B_{12}} +
%      2\,{B_{44}} -
%      {\sqrt{{{B_{11}}}^2 -
%          2\,{B_{11}}\,{B_{12}} +
%          {{B_{12}}}^2 + 16\,{{B_{14}}}^2 -
%          4\,{B_{11}}\,{B_{44}} +
%          4\,{B_{12}}\,{B_{44}} +
%          4\,{{B_{44}}}^2}}}{2} \\
% E_{44} (1,2) & = &
%   \frac{{B_{11}} - {B_{12}} + 2\,{B_{44}} +
%      {\sqrt{{{B_{11}}}^2 -
%          2\,{B_{11}}\,{B_{12}} +
%          {{B_{12}}}^2 + 16\,{{B_{14}}}^2 -
%          4\,{B_{11}}\,{B_{44}} +
%          4\,{B_{12}}\,{B_{44}} +
%          4\,{{B_{44}}}^2}}}{2} \\
% E_{44} (3) & = &
%   \frac{{B_{11}} + {B_{12}} + {B_{33}} -
%      {\sqrt{{{B_{11}}}^2 +
%          2\,{B_{11}}\,{B_{12}} +
%          {{B_{12}}}^2 + 8\,{{B_{13}}}^2 -
%          2\,{B_{11}}\,{B_{33}} -
%          2\,{B_{12}}\,{B_{33}} +
%          {{B_{33}}}^2}}}{2},
% \end{eqnarray}
% \end{widetext}
%$E_K$, $E' (1,2)$ and $E_{44} (1,2,3)$,
%in notation that refers to the corresponding moduli in the {\em bcc} phase.
%Specifically $E' (1,2)$ denotes the two eigenvalues that
%correspond to $2B'$ in the bcc phase and $E_{44} (1,2,3)$
%denotes those that correspond to $2B_{44}$ in bcc.
%and $E_K$ denotes the eigenvalue associated with three times the
%bulk modulus in bcc.

The eigenvalues are plotted in Fig.~\ref{fig:eConst}(b).
%The top curve corresponds to roughly
%three times the bulk modulus of the ground state up to 3.15~Mbar.
%and shows nearly perfect agreement with the bulk modulus of the
%bcc phase as mentioned earlier.
%
The top curve represents two degenerate eigenvalues that are equal
to $B'=(B_{11}-B_{12})/2$ in the bcc phase, the usual shear modulus for
tetragonal shear in the cubic crystal.  It is quite smooth.
The remaining three eigenvalues are degenerate in the bcc phase
and equal to $B_{44}$, the shear modulus
for trigonal shear in the cubic crystal (not to be confused with
the $B_{44}$ in the rhombohedral frame).  In the
rhombohedral phases two of these eigenvalues remain degenerate but
one splits off.  That single eigenvalue represents a pure shear
corresponding to the rhombohedral deformation.  Its value is
%\begin{equation}
%\frac{1}{3} \left( B11+2B33+B12-4B13 \right)
%\end{equation}
$( B_{11}+2B_{33}+B_{12}-4B_{13})/6$
which decreases toward zero as the pressure in the rhombohedral phases
approaches the bcc phase boundary.  This decrease is most
pronounced approaching the high-pressure reentrant bcc phase (2.8~Mbar),
but it is present at both.  In the energy curves, it is clear
that the width of the rhombohedral well is broadening with the
change in pressure as it rises above the bcc well and quickly
becomes unstable.
By the same token, the single eigenvalue reaches its maximum at 1.87~Mbar,
the pressure that the rhombohedral well is deepest and most stable
against the bcc phase.
% At the $\beta$--$\gamma$ boundary,
% it is the other $B_{44}$-like moduli, the paired moduli, that
% are approaching zero, although interestingly the $\beta$ structure
% remains metastable with an order parameter that goes smoothly
% to zero at $\sim$2.7~Mbar.\cite{Lee}
%
The eigenvalues can also be used to study the elastic anisotropy
of the crystal. The anisotropy ratio ($A=B_{44}/B'$ in bcc)
has been calculated for all four phases from the ratio of the
smallest and largest eigenvalues.
For an isotropic material $A=1$; for vanadium $1/A$ is never less than
its ambient value, fluctuates throughout the entire pressure range
studied, and becomes extremely high near the $\gamma$--bcc boundary
($A\sim 1/130$).
For comparison, the most anisotropic cubic transition metal at ambient
conditions is copper\cite{HirthLothe} with $A=3.21$, and among all
cubic elements recent calculations found for polonium
$A$=1/6 to 1/18 at T=0K.\cite{Legut}
% (expressed as $A$ since for Cu $C_{44}>C'$).

%Surprisingly, this modal analysis deconvolutes the anomalous shear
%response of the $B_{33}$ and $B_{44}$ from the dilatory response
%as can be seen from the similarity between $B_{33}$ and $E_{44} (3)$,
%and between $B_{44}$ and $E_{44} (1,2)$ (RER: what does
%this sentence mean?).
%(BL: I'm trying to make a direct connection between the raw moduli and eigenvalues so that we can better understand %the raw moduli. Anyways,
%I wanted to say, if you compare $B_{33}$ with $E_{44} (3)$,
%you can see they look almost the same except that $B_{33}$ has a higher slope.
%I think the modal analysis can signify $E_{44} (3)$-like response out of the net $B_{33}$ curve, which consists of %the bulk modulus contribution and the shear contribution.
%Hope this clarifies.)

%The similarity between $B_{33}$ and $E_{44} (3)$ confirms that $E_{44} (3)$ captures the anomalous shear response of $B_{33}$, which is of great interest here.
%$E_{44} (3)$ of the rhombohedral phase, hence $B_{33}$, increases until
%1.87~Mbar, at which the rhombohedral phase attains its greatest
%stability against the bcc phase and the largest stiffness against shear
%deformation. It decreases as the protruded rhombohedral phase in the
%energy landscape flattens back and disappears with pressure further increased.

\section{\label{sec:level5}Polycrystalline shear moduli}

\begin{figure}
\includegraphics[width=0.43\textwidth]{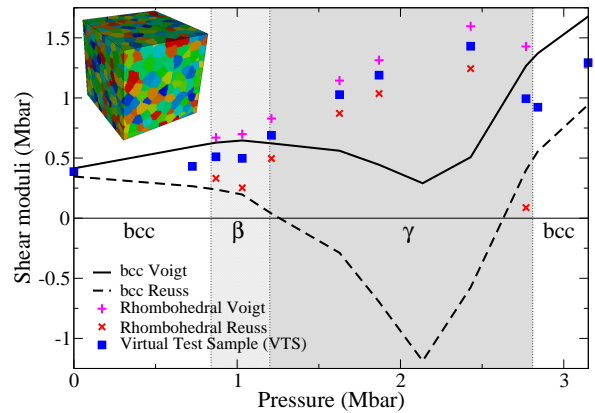}
\caption{(Color online) The polycrystalline shear modulus
as a function of pressure,
along with the Voigt and Reuss bounds for bcc, $\beta$ and $\gamma$.
This shear modulus is based on the stress-strain coefficients
$B_{ijkl}$ of the single-crystal structure
%--bcc, $\beta$, $\gamma$, and reentrant bcc
(see Fig.~\ref{fig:eConst}~(a)),
and calculated from the virtual test sample with random grain orientation shown in color (inset).
\label{fig:shearMod}}
\end{figure}

Polycrystalline vanadium without texture has isotropic
mechanical behavior, described by just two independent
elastic moduli: the bulk modulus $K$ and the shear modulus $G$.
Regardless of phase, $K$ is within 3\% of that of the bcc structure as mentioned earlier.
Using the single-crystal $B_{ijkl}$,
$G$ may be bounded by the Voigt and Reuss
approximations of constant strain and constant stress,
respectively. We have calculated these approximations using
expressions equivalent to those in the literature.\cite{Watt}
Since dynamic experiments conducted at Z-pinch and laser
facilities often use thin-film targets with microstructures
that can vary from columnar to equiaxed depending on how they
are processed, the Voigt and Reuss bounds are helpful
in assessing the range of possible responses.
In calculating the Voigt and Reuss bounds shown
in Fig.~\ref{fig:shearMod} (as well as
the explicit polycrystalline calculations below), we assume that
the deformations are infinitessimal.  With the low energy
barriers, switching between variants of the rhombohedral phase
may contribute to the strain with no cost in stored elastic
energy, leading to a reduction in the shear modulus.  At larger
strains the response to rhombohedral strains stiffens anharmonically.
Both of these effects have been neglected.
The homogenized shear modulus in the rhombohedral phase is
positive, indicating mechanical stability.
The variation in the Voigt-Reuss
difference results from the changing crystalline anisotropy.

% FIG

In the case of a microstructure with more equiaxed grains,
we calculate the polycrystalline shear
modulus by homogenizing the single-crystal $B_{ijkl}$
of the ground state structure at each
pressure using a virtual test sample (VTS).\cite{Becker}
%The finite element VTS is a cube of size (300 $\mu$m)$^3$ with 1000 randomly
%oriented grains.
The procedure in Ref.~\onlinecite{Becker} has been repeated: the VTS is strained in six pure shear modes
($\varepsilon_{12}$, $\varepsilon_{23}$, $\varepsilon_{31}$,
$\varepsilon_{11}-\varepsilon_{22}$, $\varepsilon_{22}-\varepsilon_{33}$,
and $\varepsilon_{33}-\varepsilon_{11}$), and the results from six
calculations are averaged to reduce the effects of anisotropy.
The isotropic average shear modulus($G_{VTS}$) plotted in
Fig.\ \ref{fig:shearMod} has been obtained by equating the calculated
elastic energy per volume to an ideal elastic solid at the same strain:
\begin{equation}
u= G_{VTS}(\varepsilon_{11}^2+\varepsilon_{22}^2+\varepsilon_{33}^2+\varepsilon_{12}^2+
\varepsilon_{23}^2+\varepsilon_{31}^2)\;.
\label{eq:Gvts}
\end{equation}
Only one or two of the six strains are nonzero at each run depending on which of the six modes
has been applied, and the resulting variance in six independent runs is indicative of the anisotropy.
The overall VTS prediction lies between the Voigt
and Reuss bounds, and the VTS values are within 5\% of the
Voigt-Reuss-Hill average\cite{VRH} except at the reentrant
bcc phase boundary, where the VTS value is 31\% greater.
%but it gives a better value for the shear modulus
%especially in the region just prior to
%the reentrant bcc transition, where
The Voigt-Reuss difference at this phase boundary is quite large: 1.34~Mbar.
The constant-stress Reuss average
is sensitive to the most compliant orientation, whereas the Voigt average
is fairly insensitive;  the VTS shear modulus is closer to the Voigt
value.  It significantly decreases at this point of high anisotropy,
and may lead to an anomalous dynamic response.

\section{\label{sec:level6}Conclusion}

We have investigated the properties relevant to dynamic
experiments for two high-pressure rhombohedral phases in vanadium metal.
% The first phase transformation was suggested by the softening of the C$_{44}$
% elastic constant with pressure in vanadium \cite{Suzuki,Landa}
% and confirmed experimentally in
% recent high-pressure DAC experiments.\cite{Mao}
% The rhombohedral phase has been confirmed by DFT calculations,\cite{Lee}
% where it was found that a second rhombohedral phase should
% exist at pressures above about 1.2~Mbar before encountering
% a reentrant bcc phase at about 2.8~Mbar.  All of these
% pressures are accessible to dynamic experiments.
%
It will be challenging for dynamic experiments to detect the
rhombohedral phase unambiguously.  The distortion is probably
too small for in-situ x-ray diffraction, although it might
be large enough in $\gamma$.\cite{Lee}  We have predicted that
%the volume change of
%the initial rhombohedral phase (0.3\%) is too small to have
%a clear effect, but again $\gamma$ is promising with a
%volume change for $\beta$ to $\gamma$ of 1.5\%
%that may present a signature in the VISAR trace.
the volume change associated with any phase transformation up to 
3.15~Mbar is small, and may not have a clear signature in the VISAR trace.
We have also predicted values for the single crystal and 
polycrystalline stress-strain coefficients in the rhombohedral
phases at zero temperature.
The $\beta$ and $\gamma$ phases smoothly cut off the negative
values of the bcc $B_{44}$.
The first order transitions between the bcc and rhombohedral
phases give remarkably small changes in the stress-strain coefficients,
as evident from the plots of the shear modulus and the stress-strain
matrix eigenvalues,
apart from near the $\gamma$--bcc transition where the crystal
is highly anisotropic.
% This fact is evident in both the homogenized shear modulus
% and the eigenvalues of the elastic constant matrix.  The one
% notable exception is the softening of one of the eigenvalues
% at pressures just below the $\gamma$--bcc transition, and the
% corresponding dip in the VTS and Reuss shear moduli.  The
% single-crystal elastic constants become extremely anisotropic
% at this point.

The results here were obtained using DFT at zero temperature
for pure vanadium.  Since the phase transition is driven by
rather subtle electronic structure effects, the elastic
constants may be substantially affected by changes in temperature
or impurities.\cite{Landa,Lee,ShaCohen}  It would be interesting to
see whether the remarkable continuity of the moduli persists
as the phase boundaries and the relative stiffness of the
bcc and rhombohedral structures change.

\begin{acknowledgments}
We would like to thank G.\ Collins, A.\ Landa,
D.\ Orlikowski, B.\ Remington, and
P.\ S\"{o}derlind for useful discussions.
This work was performed under the auspices of the U.S.\ Dept.\ of
Energy by Lawrence Livermore National Laboratory under
Contract DE-AC52-07NA27344.
\end{acknowledgments}

\appendix
\section{Calculation of stress-strain coefficients in the rhombohedral lattice}

\begin{table*}
\caption{\label{tab:fitting}Deformation gradients for the six independent 
stress-strain coefficients in the rhombohedral lattice and the 
corresponding strain energy relations per unit volume at pressure $P$.}
\begin{ruledtabular}
\begin{tabular}{ccc}
Stress-strain coefficient & Deformation gradient & Strain energy\\
\hline
$B_{11}$  &
$T(\delta)=\left(
\begin{array}{ccc}
1+\delta & 0 & 0\\
0 & 1 &0\\
0 & 0 & 1\\
\end{array}
\right)$ &
$u(\delta, P)= -P\delta + \frac{1}{2}B_{11}{\delta}^2$\\
%%%%%%%%%%%%%%%%%%%%%%
$B_{33}$ &
$T(\delta)=\left(
\begin{array}{ccc}
1 & 0 & 0\\
0 & 1 &0\\
0 & 0 & 1+\delta\\
\end{array}
\right)$ &
$u(\delta, P)= -P\delta + \frac{1}{2}B_{33}{\delta}^2$\\
%%%%%%%%%%%%%%%%%%%%%%
$B_{12}$ (and $B_{66}$) &
$T(\delta)=\left(
\begin{array}{ccc}
1+\delta & 0 & 0\\
0 & 1-\delta & 0\\
0 & 0 & \frac{1}{1-{\delta}^2}\\
\end{array}
\right)$ &
$u(\delta)= (B_{11}-B_{12}){\delta}^2=2B_{66}{\delta}^2$\\
%%%%%%%%%%%%%%%%%%%%%%
$B_{13}$ &
$T(\delta)=\left(
\begin{array}{ccc}
1+\delta & 0 & 0\\
0 & 1+\delta & 0\\
0 & 0 & 1+\delta\\
\end{array}
\right)$ &
$u(\delta, P)= -3P(\delta+\delta^2)+\frac{1}{2}(2B_{11}+B_{33}+2B_{12}+4B_{13}){\delta}^2$\\
%%%%%%%%%%%%%%%%%%%%%%
$B_{44}$ &
$T(\delta)=\left(
\begin{array}{ccc}
\frac{1}{1-{\delta}^2} & 0 & 0\\
0 & 1 & \delta\\
0 & \delta & 1\\
\end{array}
\right)$ &
$u(\delta)= 2B_{44}{\delta}^2$\\
%%%%%%%%%%%%%%%%%%%%%%
$B_{24}$ ($-B_{14}$) &
$T(\delta)=\left(
\begin{array}{ccc}
1+\delta & 0 & 0\\
0 & 1-\delta & \delta\\
0 & 0 & \frac{1}{1-{\delta}^2}\\
\end{array}
\right)$ &
$u(\delta)= \frac{1}{2}(2B_{11}-2B_{12}+B_{44}-4B_{24}){\delta}^2$
\end{tabular}
\end{ruledtabular}
\end{table*}
\begin{table*}
\caption{\label{tab:elasticConstants}Calculated stress-strain coefficients 
and various polycrystalline shear modulus predictions for the stable phase. 
All quantities are in units of Mbar except that volume has been scaled 
by the ambient volume $V_o$=13.518 \AA$^3$.}
\begin{ruledtabular}
\begin{tabular}{ccccccccccccc}
Stable & Volume & Pressure & Bulk &  \multicolumn{6}{c}{Single crystal stress-strain coefficients}  &  \multicolumn{3}{c}{Polycrystalline} \\
phase & ($V/V_o$) & ($P$) & modulus ($K$) & $B_{11}$ & $B_{33}$ & $B_{12}$ & $B_{13}$ & $B_{44}$ & $B_{24}$ & VTS & Voigt & Reuss \\
\hline
bcc & 1 & 0.00 & 1.82 & 2.30 & 2.18 & 1.52 & 1.64 & 0.51 & 0.18 & 0.39 & 0.41 & 0.35 \\
bcc & 0.779 & 0.73 & 4.21 & 4.79 & 4.44 & 3.74 & 4.09 & 0.87 & 0.49 & 0.43 & 0.59 & 0.27 \\
\hline
$\beta$ & 0.754 & 0.87 & 4.63 & 5.52 & 5.00 & 3.96 & 4.43 & 0.75 & 0.57 & 0.51 & 0.67 & 0.33 \\
$\beta$ & 0.729 & 1.03 & 5.09 & 6.00 & 5.63 & 4.42 & 4.84 & 0.77 & 0.65 & 0.50 & 0.70 & 0.25 \\
\hline
$\gamma$ & 0.705 & 1.20 & 5.59 & 6.17 & 6.53 & 5.16 & 5.27 & 1.29 & 0.46 & 0.69 & 0.83 & 0.50 \\
$\gamma$ & 0.659 & 1.62 & 6.77 & 7.87 & 7.89 & 6.19 & 6.23 & 1.61 & 0.52 & 1.03 & 1.14 & 0.87 \\
$\gamma$ & 0.636 & 1.87 & 7.45 & 8.79 & 8.58 & 6.73 & 6.85 & 1.81 & 0.57 & 1.19 & 1.31 & 1.04 \\
$\gamma$ & 0.593 & 2.45 & 9.02 & 10.81 & 9.85 & 8.11 & 8.38 & 2.21 & 0.70 & 1.43 & 1.60 & 1.24 \\
$\gamma$ & 0.572 & 2.78 & 10.09 & 11.56 & 9.72 & 9.03 & 9.97 & 2.29 & 0.91 & 0.99 & 1.43 & 0.09 \\
\hline
bcc & 0.568 & 2.84 & 10.15 & 11.48 & 10.64 & 9.07 & 9.91 & 2.04 & 1.19 & 0.92 & 1.37 & 0.56 \\
bcc & 0.551 & 3.15 & 10.99 & 12.71 & 11.84 & 9.70 & 10.56 & 2.37 & 1.22 & 1.29 & 1.68 & 0.94
\end{tabular}
\end{ruledtabular}
\end{table*}

The high-pressure stress-strain coefficients
$B_{ijkl}$ can be obtained in many different ways, 
but in Table~\ref{tab:fitting}, we summarize the deformation gradients and 
the corresponding strain energy relations that we used to calculate 
$B_{ijkl}$ here. The pressure term is involved in some of the strain 
energy relations, for which the deformation gradients are not volume-conserving.
The stress-strain coefficients $B_{ijkl}(P)$ 
are equal to $C_{ijkl}$ when the pressure vanishes,
as explained in detail in Chapter~2 of Ref.~\onlinecite{Wallace}.
For a recent discussion of the stress-strain coefficients $B_{ijkl}$,
see Ref.~\onlinecite{Cohen}.
For $P\ne0$, the relation is\cite{Wallace}
\begin{equation}
B_{ijkl}= -P(\delta_{jl}\delta_{ik}+\delta_{il}\delta_{jk}-\delta_{ij}\delta_{kl}) + C_{ijkl} \;.
\label{eq:BvsC}
\end{equation}
The resulting stress-strain coefficients along with the calculated 
polycrystalline shear moduli are tabulated in Table~\ref{tab:elasticConstants}.

\end{document}